\documentclass[11pt]{article}
\pdfoutput=1
\usepackage[export]{adjustbox}
\usepackage{jheppub}
\usepackage[nottoc]{tocbibind}
\usepackage[svgnames]{xcolor}
\usepackage{float}
\usepackage[utf8]{inputenc}
\usepackage{slashed,verbatim}
\pdfoutput=1
\usepackage[most]{tcolorbox}
\usepackage{epigraph,lipsum}
\usepackage{fancyhdr}
\usepackage{epigraph}
\usepackage{graphicx}
\usepackage[normalem]{ulem}

\makeatletter
\def\@fpheader{\relax}
\makeatother

\usepackage{framed}

\usepackage{lipsum}

\newcommand\blfootnote[1]{%
  \begingroup
  \renewcommand\thefootnote{}\footnote{#1}%
  \addtocounter{footnote}{-1}%
  \endgroup
}
\usepackage{titlesec}
\titleformat*{\section}{\LARGE\bfseries}
\titleformat*{\subsection}{\Large\bfseries}
\titleformat*{\subsubsection}{\large\bfseries}
\titleformat*{\paragraph}{\large\bfseries}
\titleformat*{\subparagraph}{\large\bfseries}

\usepackage[framemethod=tikz]{mdframed}

\usepackage{amsmath,epsfig}
\usepackage{amssymb,amsfonts}
\usepackage{latexsym}
\usepackage{graphicx}

\usepackage{subeqnarray}

\usepackage{graphicx}

\def\be{\begin{equation}}
\def\ee{\end{equation}}
\def\bea{\begin{eqnarray}}
\def\eea{\end{eqnarray}}

\newcommand{\bear}{\begin{eqnarray}}

\newcommand{\eear}{\end{eqnarray}}

\newcommand{\bsea}{\begin{subeqnarray}}
\newcommand{\esea}{\end{subeqnarray}}
\newbox\pippobox

\def\6{\partial}

\newcommand{\comments}[1]{}
\allowdisplaybreaks[3]

\begin{document}


\title{\centering \Huge Collective modes of polarizable holographic media in magnetic fields}
\author[\,\odot,\dagger]{Matteo Baggioli}
\author[\,\star]{, Ulf Gran}
\author[\,\star]{, Marcus Torns\"{o}}
\vspace{0.1cm}
\affiliation[\odot]{Wilczek Quantum Center, School of Physics and Astronomy, Shanghai Jiao Tong University, Shanghai 200240, China}
\affiliation{Shanghai Research Center for Quantum Sciences, Shanghai 201315.}
\affiliation[\dagger]{Instituto de Fisica Teorica UAM/CSIC,
c/ Nicolas Cabrera 13-15, Cantoblanco, 28049 Madrid, Spain}

\affiliation[\star]{Department of Physics, Division of Subatomic, High Energy and Plasma Physics, Chalmers University of Technology
SE-412 96 G\"{o}teborg, Sweden}

\emailAdd{b.matteo@sjtu.edu.cn}
\emailAdd{ulf.gran@chalmers.se}
\emailAdd{marcus.tornso@chalmers.se}

\vspace{1cm}

\abstract{We consider a neutral holographic plasma with dynamical electromagnetic interactions in a finite external magnetic field. The Coulomb interactions are introduced via mixed boundary conditions for the Maxwell gauge field. The collective modes at finite wave-vector are analyzed in detail and compared to the magneto-hydrodynamics results valid only at small magnetic fields. Surprisingly, at large magnetic field, we observe the appearance of two plasmon-like modes whose corresponding effective plasma frequency grows with the magnetic field and is not supported by any background charge density. Finally, we identify a mode collision which allows us to study the radius of convergence of the linearized hydrodynamics expansion as a function of the external magnetic field. We find that the radius of convergence in momentum space, related to the diffusive transverse electromagnetic mode, increases quadratically with the strength of the magnetic field.
}

\blfootnote{$\dagger$   \,\,\,\url{https://members.ift.uam-csic.es/matteo.baggioli}}

\maketitle

\section{Introduction}

Over the last decade holographic techniques, originating from string theory, have found a multitude of applications addressing qualitative properties of real world systems where quasiparticles are destroyed due to strong interactions/correlations \cite{Hartnoll:2016apf,Baggioli:2019rrs}. Prominent examples are the quark gluon-plasma produced by colliding heavy atomic nuclei, and the strange metal phase appearing at temperatures above the superconducting transition temperature in many materials exhibiting high temperature superconductivity. In this type of materials the holographic techniques provide access to parts of the phase space not accessible by expansion methods or, in many cases, even numerical methods. These novel techniques can therefore provide valuable insights into the physical behaviour of rather generic systems without quasiparticles, as well as generate predictions for future experiments.

When considering charged systems the dynamics generally induces a polarisation (electric and/or magnetic), which can have large effects on the physics of the system in question. Incorporating polarization effects into the holographic framework requires adding Coulomb interactions in the boundary theory, thereby making the $U(1)$ gauge field dynamical \cite{Witten:2003ya,Witten:2001ua,Mueck:2002gm}. This was recently accomplished in two different, but equivalent, ways. First by a judicious choice of boundary conditions for the bulk fields, ensuring dynamical electromagnetism in the boundary theory \cite{Aronsson:2017dgf,Aronsson:2018yhi}, and later by adding a double-trace deformation to the boundary theory \cite{Mauri:2018pzq,Krikun:2018agd}. This led to the first holographic treatments of plasmons \cite{Aronsson:2017dgf,Aronsson:2018yhi,Gran:2018vdn,electronclouds,Mauri:2018pzq,Krikun:2018agd}, i.e.~collective oscillations of the electrons, a phenomenon driven by the polarization in the medium. 

In this paper we will extend our previous studies of collective modes in polarizable holographic media \cite{Baggioli:2019aqf, Baggioli:2019sio} to incorporate a non-zero (external) magnetic field, since this is an important experimental scenario. We have chosen a simple model, a dyonic black brane in $AdS_4$, in order to elucidate the generic qualitative effects on the collective modes by turning on the magnetic field. We will also be working at zero chemical potential in order to isolate the effects from the magnetic field and the induced polarization, to get an as clear picture as possible.

For slowly varying perturbations (both in time and space) hydrodynamics is a robust framework relying mainly on conservation laws. We will start by making a detailed comparison of our results to the state-of-the-art treatment of relativistic magneto-hydrodynamics \cite{Grozdanov:2017kyl,Hernandez:2017mch,Buchbinder:2008dc,Buchbinder:2009mk,Hansen:2008tq,Huang:2011dc,Kovtun:2016lfw,Montenegro:2017rbu,Hirono:2017hsg,Hattori:2017usa,Armas:2018atq,Glorioso:2018kcp,Armas:2018zbe,Hongo:2020qpv}, i.e.~hydrodynamics coupled to dynamical electromagnetic fields \cite{Hernandez:2017mch}. 

This will allow us to get an understanding of the mechanism behind the decrease in the range of validity of hydrodynamics when the coupling strength is increased. This comparison will also clarify exactly how first order magneto-hydrodynamics starts to break down when increasing the coupling strength $\lambda$; generically first order magneto-hydrodynamics gets the point $\omega(k=0)$ on the dispersion relations correct, but the dynamics at larger momentum or frequency becomes worse approximated the larger the coupling strength is. In other words, the larger the coupling strength, the more impact the higher order corrections have in the hydrodynamic expansion.

Then, we will go beyond the regime of applicability of hydrodynamics and we will discuss the dynamics of the collective modes for large momentum, frequency and external magnetic field. Finally, we will investigate in more detail the convergence properties of the linearized hydrodynamics approximation in relation to the magnetic field strength and we will determine the critical scale at which non-hydro modes cannot be neglected anymore.

The paper is structured as follows. In section \ref{model} we present the dyonic black-brane model in $AdS_4$ we will be using. In section \ref{zeroB}, as a warm-up and for completeness, we exhibit the collective modes in the system at zero magnetic field. In section \ref{nonzeroB} we turn on a magnetic field and study how the collective modes evolve when varying the parameters at our disposal. 
Finally, in section \ref{concl} we summarize our results and discuss some avenues for future studies. 

\section{The holographic model}
\label{model}

We consider a simple Einstein-Maxwell action in four bulk dimensions defined by the action
\begin{equation}\label{eq:action}
S\,=\,\int d^4x \sqrt{-g}
\left[\frac{R}2+\frac{3}{\ell^2}\,-\,\frac{1}{4}\,F^2\right]
\end{equation}
together with the standard definition $F=dA$.

This setup admits an asymptotic AdS black-brane solution with metric
\begin{equation}
\label{backg}
ds^2=\frac{1}{z^2} \,\left[\,-f(z)\,dt^2\,+\,\frac{dz^2}{f(z)}\,+\,dx^2\,+\,dy^2\,\right]~.
\end{equation}
The holographic $z$ coordinate goes between the AdS boundary $z=0$ and the black hole horizon $z=z_h$ (defined as $f(z_h)=0$).
The profile for the bulk gauge field is chosen to be
\begin{equation}
    A\,=\,A_t(z)\,dt\,+\,\mathcal{B}\,x\,dy\,,\qquad A_t(z)\,=\,Q\,\left(1-z\right)\,.
\end{equation}
In this notation, $\mathcal{B}$ is the bulk magnetic field while the chemical potential, charge density and magnetic field of the dual theory are given by
\begin{align}
   \mu			\;=\;	\frac{Q}{\sqrt{\lambda}\, z_h}	\, , \qquad
\rho	\;=\;	\frac{\sqrt{\lambda}\, Q}{z_h^2}\,, \qquad B=\frac{\mathcal{B}}{\sqrt{\lambda}z_h^2}\, ,	
\end{align}
where $\lambda$, whose meaning will be clarified in the following, governs the strength of the Coulomb interactions in the boundary field theory.

Finally, the emblackening factor is given by
\begin{equation}\label{backf}
f(z)= 1-\left(\frac{z}{z_h}\right)^3+\frac{1}{2}\left(\mathcal{B}^2+Q^2\right)\left(\frac{z}{z_h}\right)^3\left(\frac{z}{z_h}-1\right) 
\end{equation}
and the corresponding temperature of the dual field theory is
\begin{equation}
T=-\frac{f'(z_h)}{4\pi}=\frac{6-Q^2-\mathcal{B}^2}{8\pi z_h}\,.\label{eq:temperature}
\end{equation}
The gauge field perturbations on top of this background display an asymptotic expansion close to the UV AdS boundary which takes the form
\begin{equation}
    \delta A_\mu\,=\,\delta A_\mu^{(0)}\,+\,\delta A_\mu^{(1)}\,z\,+\,\mathcal{O}(z^2)~.
\end{equation}
Crucially, in this manuscript, we use the mixed boundary conditions
\begin{align} 
   \omega^2\delta A_x^{(0)}\,+\,\lambda\,\delta A_x^{(1)}\,=\,0\,,\\
   \left( \omega^2\,-k^2\right)\delta A_y^{(0)}\,+\,\lambda\,\delta A_y^{(1)}\,=\,0\,,\label{eq:bcs}
\end{align}
which are equivalent to the familiar conditions of the dielectric function, e.g~$\epsilon(\omega,k)=0$ for longitudinal excitations \cite{nozieres1999theory}, and allow for dynamical electromagnetic interactions in the dual field theory. This is a crucial difference with respect to the previous studies considering the same Reissner-Nordström dyonic black hole solution \cite{Hartnoll:2007ih, Amoretti:2020mkp}. In the limit $\lambda=0$, the b.c.s. defined in equation~\eqref{eq:bcs} reduce to the standard Dirichlet conditions. In this sense, $\lambda$ defines the strength of the Coulomb interactions in the dual field theory.

In order to isolate the effects of the magnetic field, in this manuscript we will consider only the neutral case $Q=0$.

\section{Warm up: Zero magnetic field}
\label{zeroB}

\begin{figure}[ht]
    \centering
    \includegraphics[width=
\linewidth]{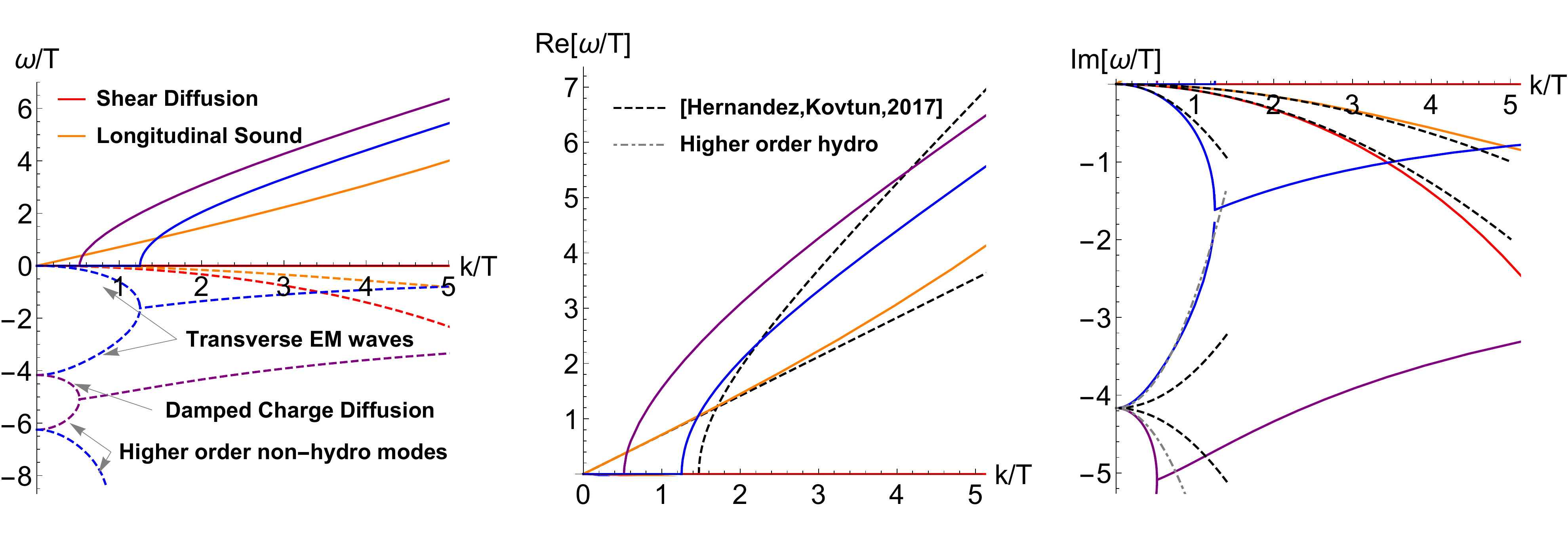}
    \caption{The dispersion relations of the collective modes in the longitudinal and transverse sectors in absence of magnetic field, $B=0$. \textbf{Left: } The full spectrum. Dashed lines indicate the imaginary parts and filled lines the real parts. The modes corresponding to each color are indicated. \textbf{Center: } Comparison between the real parts of the holographic results to those following from the hydrodynamic theory of \cite{Hernandez:2017mch} (black dashed lines). It is not surprising that the k-gap phenomenon is not well described by linearized hydrodynamics, since $k^*/T$ is $\mathcal{O}(1)$. \textbf{Right: } Comparison between the imaginary parts of the holographic results and those following from hydrodynamics \cite{Hernandez:2017mch} (black dashed lines) and a simple higher order extension (gray dotdashed lines). Again, it is not a surprise that the modes with $\omega/T,k/T \gg 1$ are not well accounted for by linearized hydrodynamics.}
    \label{fig:1}
\end{figure}

We start  by considering the scenario with zero applied magnetic field $B=0$. In this case, the dual field theory represents a neutral system with dynamical EM interactions and should be well described by a simplified version of the magneto-hydrodynamic theory of \cite{Hernandez:2017mch}. In absence of background charge density, the gravitational sector and the electromagnetic one decouple. The low-energy modes expected in this system are~\cite{Hernandez:2017mch}:
\begin{itemize}
    \item A shear diffusion mode in the transverse sector with dispersion relation
    \begin{equation}
        \omega\,=\,-\,i\,D_T\,k^2\,+\dots\,,\qquad D_T=\frac{\eta}{s\,T} \,,
    \end{equation}
    with $\eta$ and $s$ being, respectively, the shear viscosity and the entropy density of the dual field theory. Notice that in this simple case the shear viscosity saturates the Kovtun-Son-Starinets bound $\eta/s=1/4\pi$ \cite{Policastro:2001yc}.
    \item A longitudinal attenuated sound mode with dispersion relation
    \begin{equation}
        \omega\,=\,v_s\,k\,-\,\frac{i}{2}\,\Gamma\,k^2\,+\,\dots\,,\qquad v_s^2=\frac{\partial p}{\partial \epsilon}\,,\quad \,, \Gamma=\frac{\eta}{s\,T}
    \end{equation}
    where $p$ and $\epsilon$ are, respectively, the pressure and energy density of the dual field theory. Since our boundary theory is conformal, we have $\frac{\partial p}{\partial \epsilon}=1/2$.
\item A damped charge diffusion mode in the longitudinal sector with dispersion relation
\begin{equation}
    \omega=-\frac{i\,\sigma}{\epsilon_e}-i\left(\frac{\sigma}{\chi_{\rho\rho}}\right)\,k^2\,+\,\dots\label{uno}
\end{equation}
where $\sigma$ is the electric conductivity and $\chi_{\rho\rho}$ the charge susceptibility. Since we are working at zero charge density, we have $\sigma=1$ with only the incoherent contribution appearing. Notice how in the presence of EM interactions charge does not diffuse anymore but instead relaxes at a rate proportional to the EM interaction strength. 
\item A pair of transverse EM waves satisfying the linearized equation
\begin{equation}
    \omega\left(\omega+\frac{i\, \sigma}{\epsilon_e}\,+\dots\right)=\frac{k^2}{\epsilon_e\,\mu_m}+\dots\,.\label{due}
\end{equation}
This equation takes the name of telegrapher (or equivalently k-gap) equation~\cite{Baggioli:2019jcm} and at low energy it displays the following two excitations
\begin{equation}
    \omega\,=\,-\,\frac{i\, \sigma}{\epsilon_e}+\frac{i\,k^2}{\sigma\,\mu_m}\,\dots\,,\qquad  \omega\,=\,-\,\frac{i\,k^2}{\sigma\,\mu_m}\dots \label{appro}
\end{equation}
Here, we have $\epsilon_e\,\mu_m=v^2=1/2$. Again, in the absence of EM interactions, we recover from equation~\eqref{due} a pair of propagating EM waves $\omega=\pm v k$ whose speed is fixed by the requirement of conformality.
\end{itemize}
We are now in the position to compare the hydrodynamic expectations to our numerical data which are obtained to all orders in frequency and momentum. We display the dispersion relation of the collective modes for $\lambda=1$ in figure~\ref{fig:1}. For the modes coming from the gravitational sector (shear diffusion and longitudinal sound) the hydrodynamic predictions work extremely well, even beyond their regime of applicability, i.e.~up to $k/T \approx 3$. At the same time, the collective modes which belong to the gauge sector are also well described by hydrodynamics but only in the expected window $\omega/T,k/T\ll 1$. In particular, for $\lambda=1$, the only mode living in that window is the diffusive EM transverse mode. Its (damped) partner mode and the damped charge diffusion mode display large imaginary parts and therefore they are not well approximated by the simplest formulas proposed above. Nevertheless, a simple phenomenological extension of the type:
\begin{equation}
      \omega\left(\omega+\frac{i\, \sigma}{\epsilon_e}\,+\,i\,D\,k^2\right)=\frac{k^2}{\epsilon_e\,\mu_m}+\dots\,
\end{equation}
already improves substantially the agreement. A correction of this type is definitely expected once the hydrodynamic framework is pushed to higher order.

\begin{figure}[ht]
    \centering
   \includegraphics[width=0.3\linewidth]{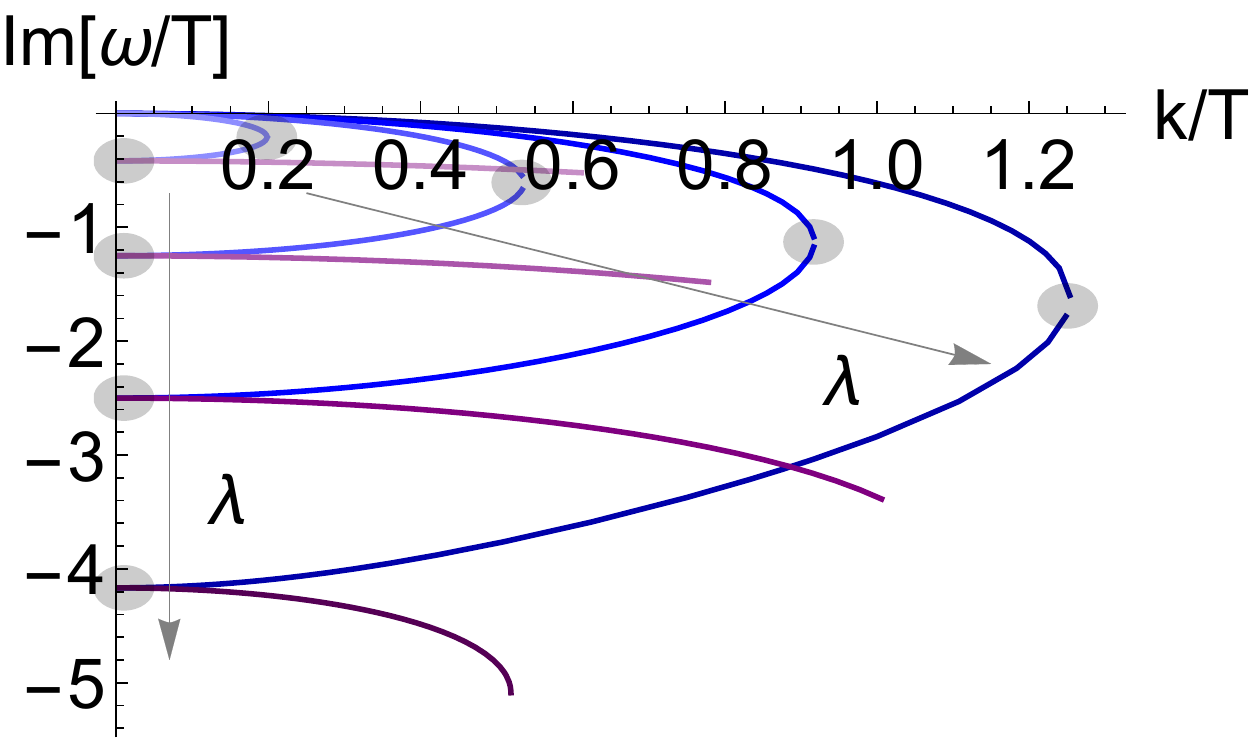}\quad
    \includegraphics[width=0.3\linewidth]{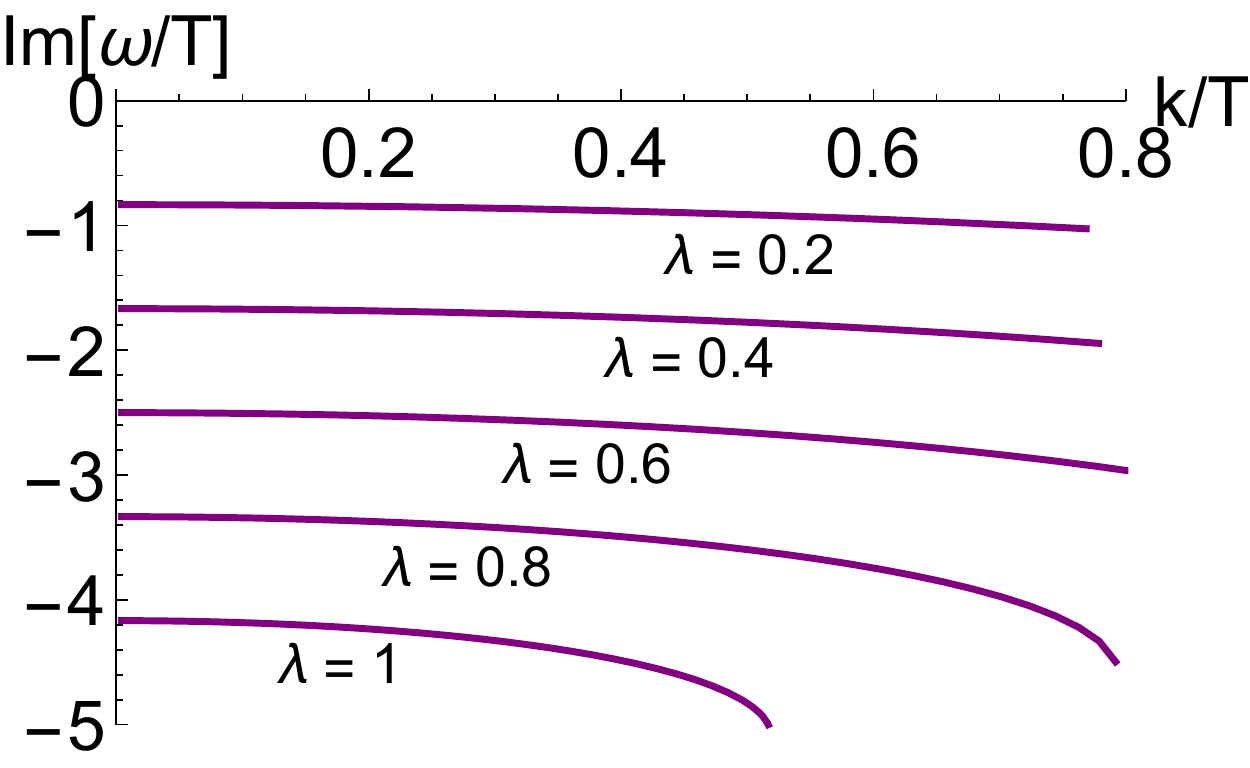}\quad
    \includegraphics[width=0.3\linewidth]{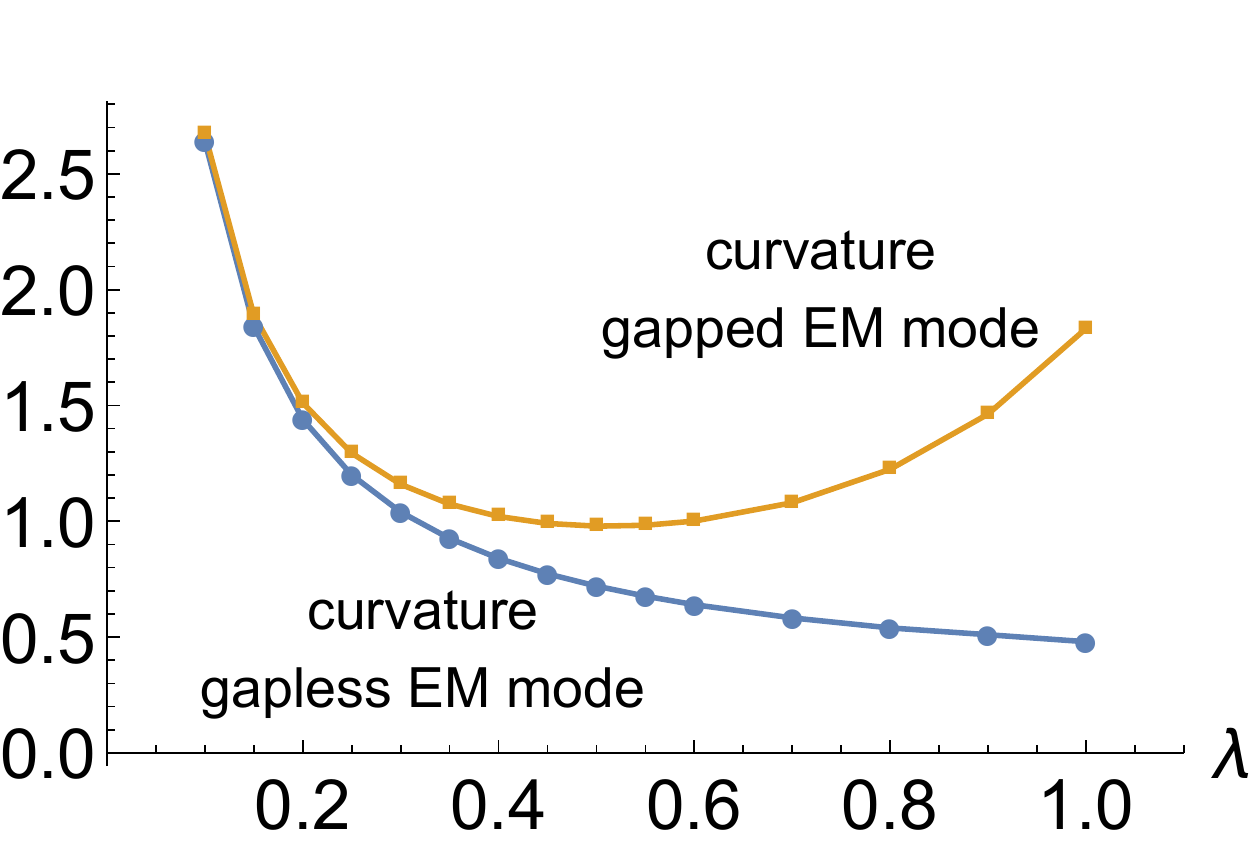}\qquad
    \caption{\textbf{Left: }The dispersion relations of the collective modes at zero magnetic field as a functions of the strength of EM interactions, $\lambda=0.1,0.3,0.6,1$. \textbf{Center: }The movement of the damped charge diffusive mode when changing the EM interactions strength. \textbf{Right: }The comparison between the curvatures (quadratic in $k$ terms) of the two transverse EM waves (blue modes in the left panel) by varying $\lambda$.}
    \label{fig:2}
\end{figure}

In figure~\ref{fig:2}, we analyze the behaviour of the collective modes upon dialing the strength of the electromagnetic interactions, our $\lambda$ parameter.
The EM interactions control (I) the damping of the charge diffusion mode (purple) and (II) the gap of the transverse EM waves (blue). In particular, it can be seen from the left panel of figure~\ref{fig:2} that both quantities grow with $\lambda$. Also, the damping of the charge diffusion mode grows linearly with $\lambda$ as expected from the hydrodynamic equation~\eqref{uno} (see center panel of figure~\ref{fig:2}). Finally, note that when diminishing the EM interactions the collective modes from the gauge sector become more long lived and move towards the hydrodynamic window. In that respect, the formulas presented above are expected to work better and better by decreasing $\lambda$. This is exactly what is observed in the right panel of figure~\ref{fig:2} by considering the curvature (the coefficient of the quadratic $k^2$ term) of the two transverse EM modes. At small $\lambda$, the curvatures approach each other as indicated by the linear order hydrodynamic analysis, see equation~\eqref{appro}.

\section{Collective modes in finite magnetic field}
\label{nonzeroB}
\begin{figure}[ht]
    \centering
    \includegraphics[width=\linewidth]{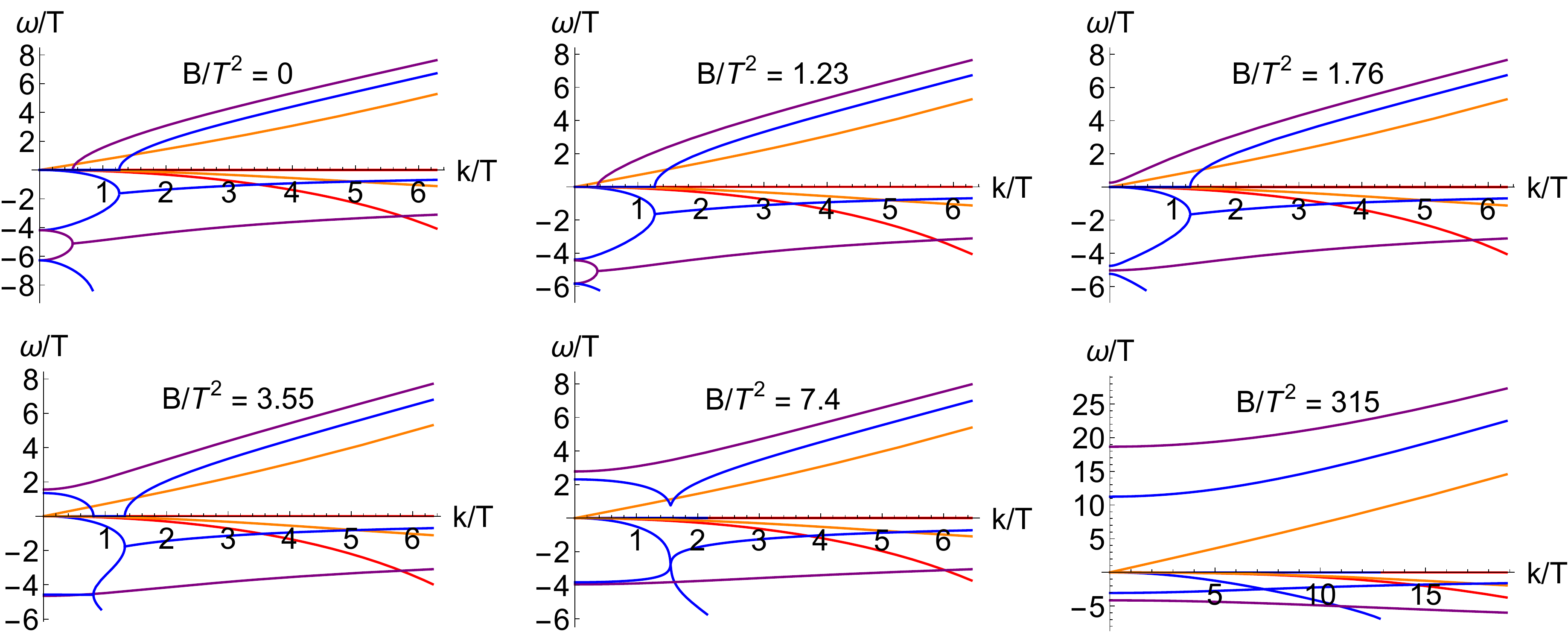}
    \caption{The dispersion relations of the collective modes at finite magnetic field. All the curves displayed in the lower half plane refer to the imaginary parts of the dispersion relations. We increase the dimensionless magnetic field strength from $B/T^2=0$ (top left panel) to $B/T^2=315$ (bottom right panel). Here the color scheme is: orange = longitudinal sound, red = shear diffusion, blue = EM transverse waves and purple = EM longitudinal waves.}
    \label{fig:3}
\end{figure}

We are now ready to switch on a finite magnetic field $B$ and study the behaviour of the collective modes for arbitrary large values of it. We display the evolution of the collective modes upon increasing the dimensionless magnetic field strength $B/T^2$ in figure~\ref{fig:3}. Notice that, at finite magnetic field, the transverse and longitudinal sectors couple. Despite this coupling induced by the magnetic field, the dynamics of the modes originating in the gravitational sector, i.e.~the longitudinal sound mode and the transverse shear diffusion mode, remains almost unaffected. In particular, from a qualitative point of view, their dispersion relations do not change. In a sense, the only effect of the finite magnetic field is to modify the transport coefficients entering in those expressions. 
The story is more interesting for the EM modes (blue and purple in figure~\ref{fig:3}). Several interesting effects appear. 
At zero magnetic field, the damped charge diffusion mode (lowest purple mode) display a k-gap shape through a collision with a higher non-hydrodynamic mode around $k_1^*/T \approx 0.56$ while the transverse EM waves (blue) collides around $k_2^*/T \approx 1.25$. 
By increasing the magnetic field, the first critical point $k_1^*/T$ moves to lower values while the second one $k_2^*/T$ moves to larger values. 
The higher k-gap shape (purple) shrinks while the lower (blue) becomes larger. 
At a certain critical magnetic field (approximately the top right panel of figure~\ref{fig:3}), the two purple modes collide already at $k=0$ and the imaginary part of the dispersion relation of the damped charge diffusion mode becomes approximately constant in momentum. Beyond this point, the damped charge diffusion mode acquires a finite real value at zero momentum. 
This implies the spontaneous appearance of an effective mass, or plasma frequency, for such mode which is exclusively produced by the magnetic field $B$. Notice that this happens for $B/T^2 \approx 1.5$ and therefore definitely outside the regime of applicability of standard magneto-hydrodynamics. Increasing further the value of the magnetic field, also the transverse EM waves start to display a non-trivial dynamics which involves three different modes. This dynamics is reminiscent of what already observed in \cite{Baggioli:2019aqf,Baggioli:2019sio}, but now at exactly zero charge density. Finally, for very large magnetic fields, $B/T^2 \,\gtrsim 10$ both the EM waves display an emergent, and different, plasma frequency. Their real part is very well approximated by a dispersion of the type
\begin{equation}
    \mathrm{Re}\,\omega^2\,=\, \omega_p^2\,+\,a\,k^2\,+\dots
\end{equation}
where, by analogy, we denote the mass terms by $\omega_p$ -- the effective plasma frequencies.
In this final regime, the imaginary parts of the electromagnetic modes become much smaller than the corresponding real part, and hence these modes become more long-lived when turning on very large magnetic fields. Note however that the modes are still damped even at $k=0$.

A signature of these modes could potentially be observable in the response functions of a strongly correlated system at large magnetic field.
\\
Now, let us focus in more detail on the various collective modes and their properties.

The first mode we investigate is the shear diffusion mode, shown in red color in figures \ref{fig:1} and \ref{fig:3}, and displaying a dispersion relation of the type
\begin{equation}
    \omega\,=\,-\,i\,D_T\,k^2\,+\,\dots \,,
\end{equation}
where $D_T$ is the shear diffusion constant. At leading order in $B/T^2$ this diffusion constant is given by $D_T=\eta/(\epsilon+p)$ and at $B=0$ satisfies exactly $D_T T=1/4\pi$, the famous KSS value \cite{Policastro:2001yc}.
\begin{figure}
    \centering
    \includegraphics[width=0.45 \linewidth]{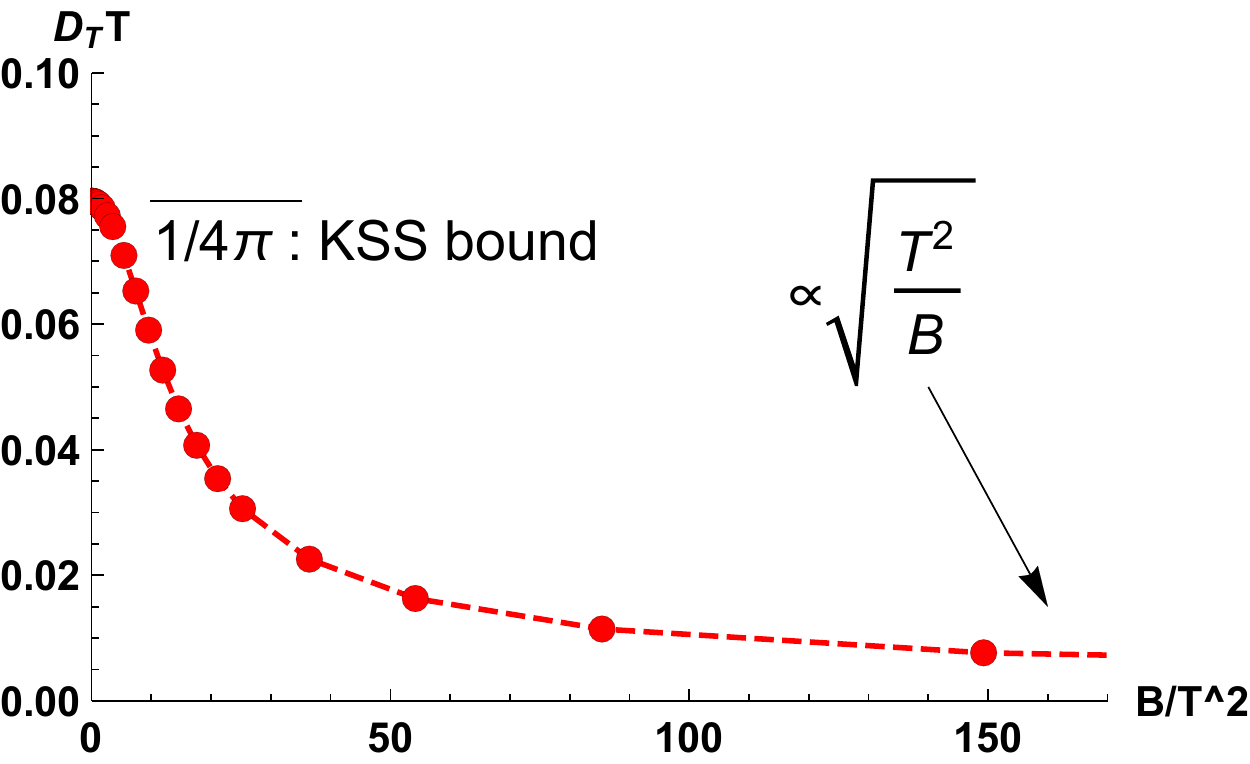}
    \qquad
    \includegraphics[width=0.45 \linewidth]{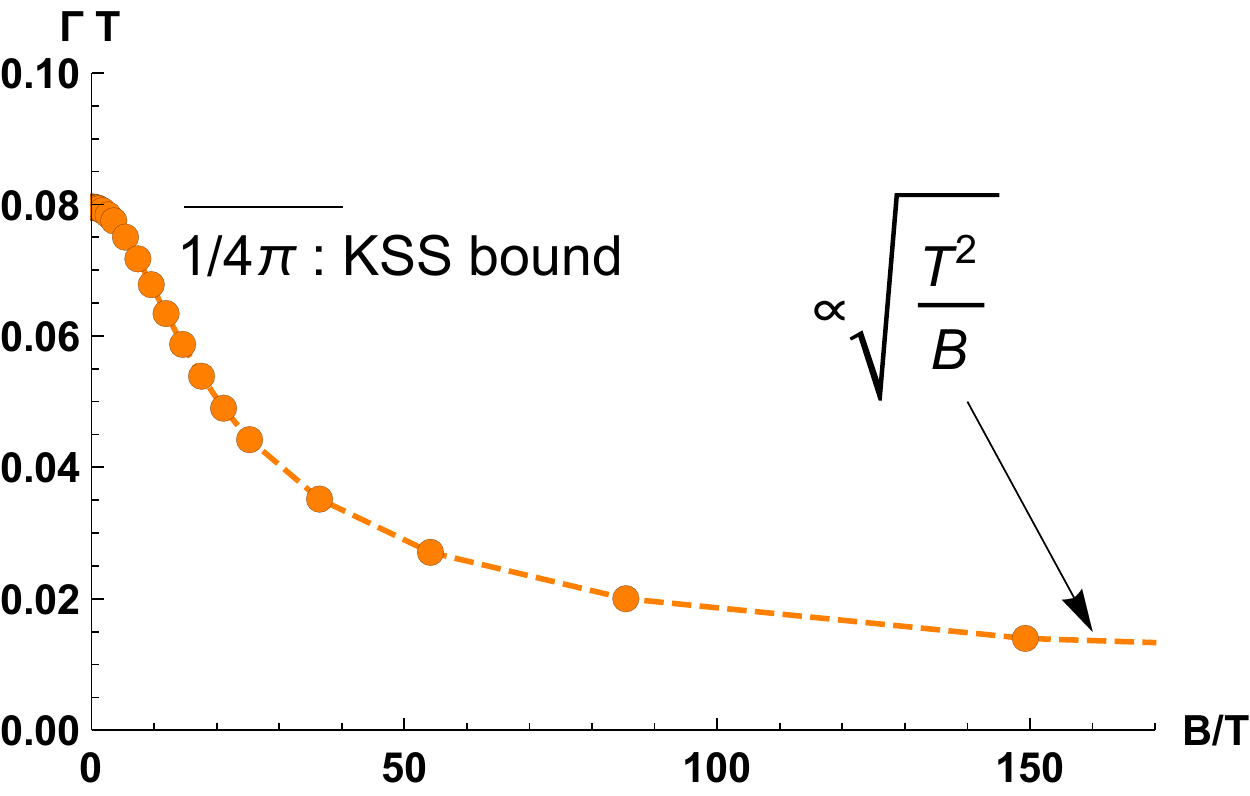}
    \caption{The dimensionless shear diffusion constant and longitudinal sound attenuation rate as functions of the magnetic field strength.}
    \label{fig:4}
\end{figure}
Increasing the magnetic field, the value of the dimensionless diffusion constant goes below the KSS value and at large magnetic field it drops to zero with a power-law scaling $D_T T \sim \sqrt{T^2/B}$ (see left panel of figure~\ref{fig:4}). This dynamics is similar to what is observed in anisotropic systems with external magnetic fields or axions sources \cite{Alberte:2016xja,Jain:2014vka,Rebhan:2011vd,Finazzo:2016mhm,Ammon:2020rvg}. 
Notice however a difference with respect to the anisotropic backgrounds.
There, the KSS bound is violated only in the direction parallel to the anisotropy (the magnetic field or the axion source); here, in two spatial dimensions, the magnetic field is a scalar, the background is still isotropic and the KSS bound is violated in all directions.

Let us move to the longitudinal sound mode (orange curves in figures \ref{fig:1} and \ref{fig:3}), whose dispersion relation is given by
\begin{equation}
    \omega\,=\,v_s\,k\,-\,i\,\frac{\Gamma}{2}\,k^2\,+\,\dots \,,
\end{equation}
where $v_s$ is the speed of sound and $\Gamma$ the sound attenuation constant. Not surprisingly, because of conformal symmetry, the speed of longitudinal sound takes the standard value $v_s^2=1/2$. In contrast, the attenuation rate does depend on the magnetic field strength. At small magnetic field, the attenuation constant is given by $\Gamma=\eta/(\epsilon+p)$ and at zero magnetic field it again takes the universal KSS value $\Gamma = 1/4 \pi T$. Increasing the magnetic field, the attenuation rate decreases and longitudinal sound becomes more and more long lived. For large magnetic field, $B/T^2 \gg 1$, we find that $2D_T\,=\,\Gamma$ and that both quantities scale like $\sim \sqrt{T^2/B}$ .

In the last part of this analysis, we move on to the modes coming from the electromagnetic sector (purple and blue in figures \ref{fig:1} and \ref{fig:3}). We have already explained above the more complicated dynamics displayed; here, we focus on two features: (I) the value of the critical momenta $k_1^*$ and $k_2^*$ at low magnetic field and (II) the value of the effective plasma frequencies at large magnetic field.
\begin{figure}
    \centering
    \includegraphics[width=0.45 \linewidth]{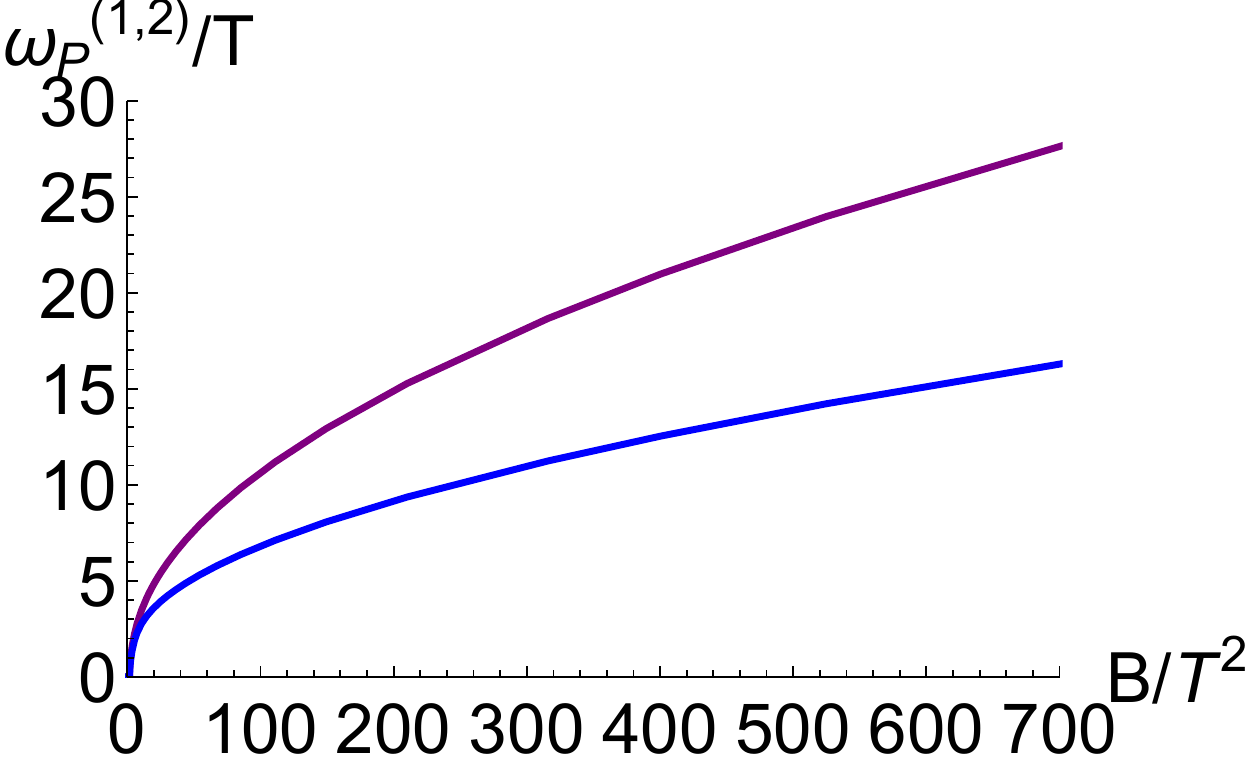}
    \qquad
     \includegraphics[width=0.45 \linewidth]{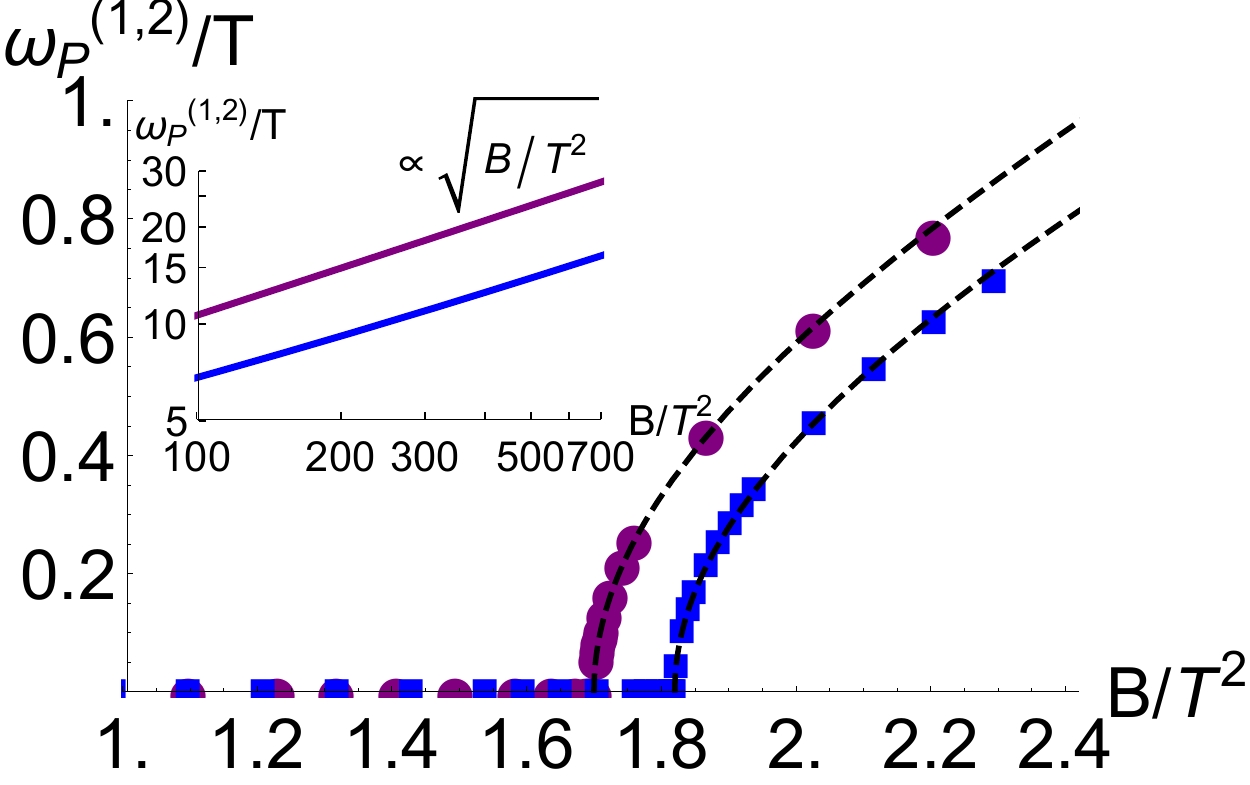}
    \caption{The frequency gap of the EM transverse mode and of the damped charge diffusion mode at large magnetic field. The right panel shows a zoom at small magnetic field and one at large magnetic field. Each of the modes has a critical magnetic field at which the frequency gap appears. Moreover, at large magnetic field, both frequency gaps scales like $\sim \sqrt{B/T^2}$.}
    \label{fig:5}
\end{figure}
We plot the frequency gaps $\omega_p^{(1,2)}$ in figure~\ref{fig:5}. The first important feature is that these frequency gaps appear only above a certain magnetic field threshold. For both modes this critical magnetic field is $\mathcal{O}(1)$ and it is slightly smaller for the damped charge diffusion mode (which is indicated with purple color and $^{(1)}$ index). Close to the critical point, the two curves are well approximated by a square root behaviour. More specifically, we find that the following functions,
\begin{align}
    &\omega_p^{(1)}\,=\,0.56\, T\,\sqrt{\frac{B^2}{T^4}-2.89}+\dots\,,\\
    &\omega_p^{(2)}\,=\,0.51\, T\,\sqrt{\frac{B^2}{T^4}-3.31}+\dots\,,
\end{align}
accurately approximate the numerical data. We note that the critical magnetic field should be easy to obtain experimentally, and the limiting factor for observing the phenomena outlined above is probably the ability of preparing a system close enough to charge neutrality, e.g.~in graphene, to avoid finite charge effects dominating the dynamics. 

At very large magnetic field, we find that both gaps scale like $\sim \sqrt{B/T^2}$. This feature is similar to what was found in a different context in \cite{Baggioli:2020edn}. This suggest that such a scaling is just a property of the UV fixed point and that it might be possible to derive it by dimensional analysis.

Finally, we are interested in studying the dynamics of the critical momentum scales $k^*_1$ and $k^*_2$. The second one, corresponding to the transverse EM modes, is particularly interesting because it locates the collision between a diffusive EM mode and a second non-hydrodynamic mode. This collision, and in particular the momentum $k^*_2$ at which it happens, determines the radius of convergence of the linearized hydrodynamic expansion in this sector. More precisely, as shown in \cite{PhysRevLett.122.251601,Grozdanov2019,Withers:2018srf} and studied further in \cite{Abbasi:2020ykq,Jansen:2020hfd,Baggioli:2020loj,Arean:2020eus,weijiaandi}, the radius of convergence in momentum space, $\mathcal{R}$, is given by
\begin{equation}
    \mathcal{R}\,=\,|k_2^*|\,.\label{hy}
\end{equation}
To the best of our knowledge, no study of the radius of convergence of hydrodynamics as a function of the external magnetic field $B$ has been reported so far. As an important disclaimer, let us emphasize that our analysis is restricted to the transverse diffusive EM mode. In particular, we cannot exclude an earlier breakdown of the linearized hydrodynamic series due to the collision (this time in the complex momentum plane) of other modes.

\begin{figure}
    \centering
    \includegraphics[width=0.45 \linewidth]{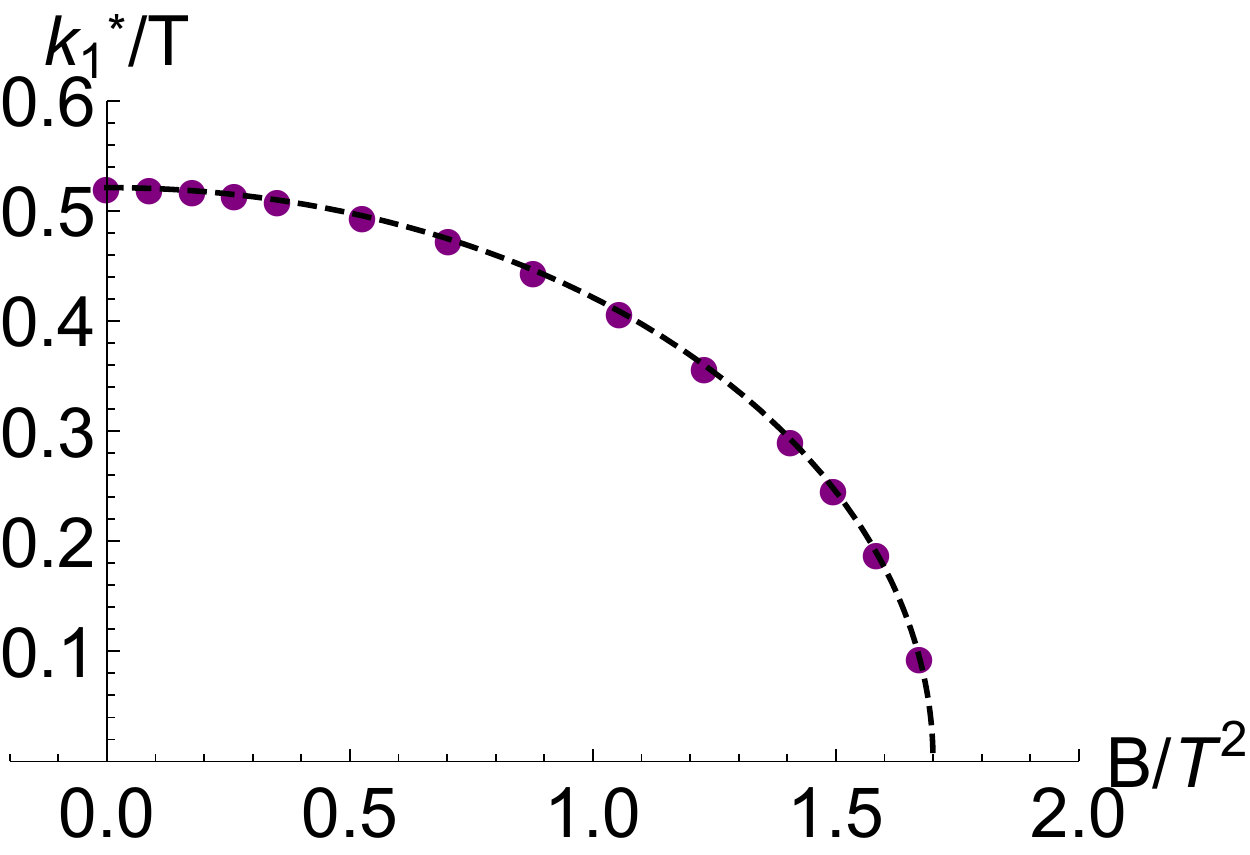}
    \qquad
     \includegraphics[width=0.47 \linewidth]{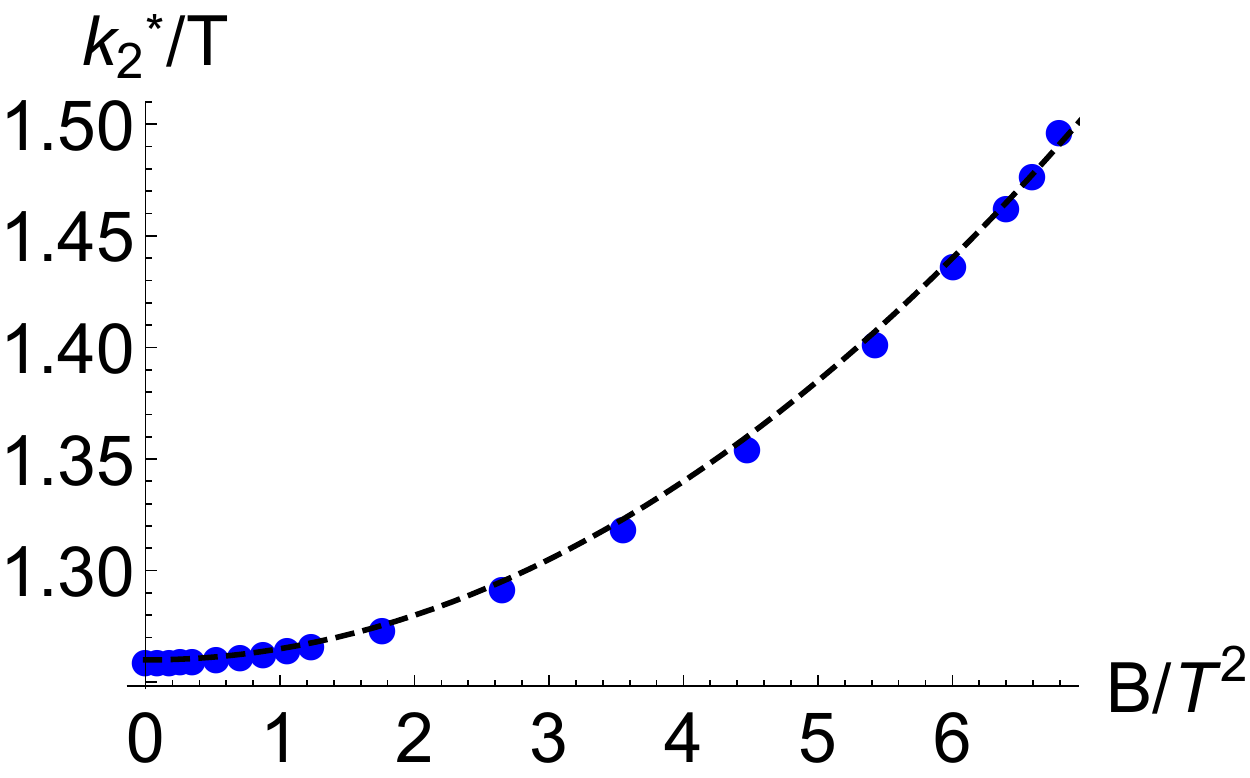}
    \caption{ \textbf{Left: }The critical momentum $k_1^*$ coinciding with the collision of the two purple modes, one of them being the damped charge diffusion mode. \textbf{Right: } The critical momentum $k_2^*$ at which the diffusive transverse EM mode collides with a first non-hydro mode. This parameter determines the radius of convergence of the hydrodynamic expansion as in equation~\eqref{hy}.}
    \label{fig:6}
\end{figure}

We show our results in figure~\ref{fig:6}. We start by studying the critical momentum of the damped charge diffusion mode in the left panel of figure~\ref{fig:6}. The data are well fitted by the simple function:
\begin{equation}
    k_1^*\,=\,0.3\,T\,\sqrt{1.89-\frac{B^2}{T^4}}\label{ff1}\,.
\end{equation}
The collision between the two purple modes disappear after a critical value of the magnetic field, approximately $B/T^2\sim\,1.7$, i.e.~the mode acquires a real part for all momenta.

We now move to the second critical point $k_2^*$ which is determined from the collision of the transverse diffusive EM mode and the first non-hydrodynamic mode (both displayed in blue in the figures). We show the value of the normalized momentum as a function of the dimensionless magnetic field in the right panel of figure~\ref{fig:6}. We observe a monotonic increase which is well-fitted by the following function
\begin{equation}
    k_2^*\,=\,1.26\,T\,+\,0.005\,\frac{B^2}{T^3}\label{ff2}\,.
\end{equation}
At a critical value of the magnetic field, around $ B/T^2 \sim 7.4$, the collision disappears from the dynamics of the quasinormal modes. This implies that beyond such point the radius of convergence of hydrodynamics cannot be defined anymore by looking solely to real values of the momentum $k$ and a more complicated analysis is needed\footnote{From observation of the fifth panel of figure \ref{fig:3}, at large values of $B/T^2$, the critical momentum is not anymore real and in the dispersion relation it appears to be replaced by an inflection point in the imaginary part.}.

Before that point, which is far beyond the small magnetic field regime $B/T^2 \ll 1$, we notice that the radius of convergence of hydrodynamics in momentum space grows with the magnetic field. In other words, the linearized hydrodynamic expansion for the transverse diffusive EM mode breaks down at larger distances for larger magnetic field. To the best of our knowledge, this is the first time this observation has been made and supported by concrete numerical data. The behaviour as a function of the magnetic field is similar to that for small charge density reported in \cite{Jansen:2020hfd,Abbasi:2020ykq}.

\section{Conclusions}
\label{concl}
In this work, we have performed a careful analysis of the collective modes of a neutral holographic plasma in presence of an external magnetic field $B$. We have compared our numerical results at small magnetic field to the existing magneto-hydrodynamics framework and extended the latter to large values of the magnetic field $B/T^2\gg 1$.

The most interesting outcome of this analysis is the presence of gapped modes at zero charge density, whose frequency depends exclusively on the magnetic field strength. Despite the similarity with magnetoplasmon \cite{Baggioli:2020edn} modes and cyclotron modes \cite{Hartnoll:2007ih}, none of them, at least in their standard incarnation, could exist at zero charge density. We suspect that the existence of such modes is supported by the presence of an incoherent conductivity in our system. Along the lines of \cite{Grozdanov:2016tdf,Grozdanov:2017kyl}, it would be interesting to investigate further the nature of these modes and their possible experimental observation. 

As expected in systems with dynamical Coulomb interactions, electromagnetic waves do not propagate anymore at large distances but rather diffuse. In contrast, at short distances, screening is not effective and propagating transverse EM waves reappear. This phenomenon is observed via a collision between the diffusive EM mode and a second non-hydrodynamic mode which takes place for a real value of the momentum $k$ and is therefore visible in the spectrum of collective modes. Exploiting this dynamics, we have been able for the first time to determine the radius of convergence of linearized hydrodynamics as a function of an external magnetic field $B$. We find that the critical momentum increases quadratically with the strength of the magnetic field.

An obvious extension of our work is to introduce a charge density background $Q\neq 0$. Here the incoherent Hall conductivity, studied in \cite{Amoretti:2020mkp} for a non-dynamical external magnetic field, is expected to be needed for matching the holographic results to magneto-hydrodynamics. 
A second, and perhaps more interesting, continuation of our program is to consider the same setup in one additional dimension, allowing for an arbitrary angle $\theta$ between the magnetic field $B$ and the wave-vector $k$. Finally, it would be important to better understand the relationship between our modified boundary conditions and the higher-forms bulk theories \cite{Grozdanov:2017kyl}.

We plan to come back to these questions in the near future.\\

\noindent \textbf{Note Added}\\
While our manuscript was in preparation, reference~\cite{kiselev2021universal} appeared with a parallel hydrodynamic treatment of  two dimensional Coulomb
interacting liquids. Qualitatively, where they overlap, our results seem in good agreement with what is observed therein. 

\section*{Acknowledgements}
We thank Egor Kiselev for useful comments and for explaining us the results of Ref.\cite{kiselev2021universal}.
M.B. acknowledges the support of the Shanghai Municipal Science and Technology Major Project (Grant No.2019SHZDZX01) and of the Spanish MINECO “Centro de Excelencia Severo Ochoa” Programme under grant SEV-2012-0249. U.G.~and M.T.~are supported by the Swedish Research Council.

\bibliographystyle{JHEP}
\bibliography{plas2}
\end{document}